\documentclass[aps,showpacs,amsmath,amssymb,nofootinbib,preprintnumbers]{revtex4}

\usepackage{graphicx}
\usepackage{color}

\def \be  {\begin{equation}}
\def \ee  {\end{equation}}
\def \bea {\begin{eqnarray}}
\def \eea {\end{eqnarray}}

\def \Tr  {\bf{Tr}}

\begin{document}

\preprint{ECTP-2010-05}

\title{Phase Space and Dynamical Fluctuations of Kaon--to--Pion Ratios}

\author{A.~Tawfik}
\email{drtawfik@mti.edu.eg}
\affiliation{Egyptian Center for Theoretical Physics (ECTP), MTI University,
 Cairo-Egypt}

\date{\today}

\begin{abstract}
The dynamical fluctuations of kaon--to--pion ratios have been studied over a wide range of center--of--mass energies $\sqrt{s}$. Based on changing phase space volume which apparently is the consequence of phase transition from hadrons to quark--gluon plasma at large $\sqrt{s}$, single--particle distribution function $f$ is assumed to be rather modified. Varying $f$ and phase space volume are implemented in the grand--canonical partition function, especially at large $\sqrt{s}$, so that hadron resonance gas model, when taking into account the  experimental acceptance and quark phase space occupation factor, turns to be able to reproduce the dynamical fluctuations over the entire range of $\sqrt{s}$.  
\end{abstract}

%\keywords{Fluctuation phenomena in statistical physics, Particle production (relativistic collisions), Quark confinement, Statistical theories of nuclear reactions}

\maketitle

\section{Introduction}

Studying the collective properties of hot and dense hadronic matter is one of the main objects of heavy-ion physics. The dependence of transport coefficients, phase structure and effective degrees of freedom on the incident energy and system size likely provides fruitful tools to study the collective properties. Studying fluctuations in particle physics of cosmic--rays dates back to early eighties of the last century \cite{crs1}.
The phase structure and event--by--event fluctuations~\cite{shury,rajag,koch} have been suggested to provide comprehensive characteristics of the particle production yields in heavy--ion collisions. They {\it statistically} variance around the averages and their scales and {\it physically} can be related to hadro-chemical composition of the particle source \cite{ebe-ph} and therefore to the phase transition.
In this regard, examining the hypothesis about the equilibrium of chemical processes in the hadronic final state~\cite{giorg} is very much essential. The dynamical fluctuations of certain particle yields have been studied at SPS and RHIC energies~\cite{Roland,NA491,STAR}. Therefore, it is natural to study the energy-dependence of the particle yield ratios and the event--by--event fluctuations using hadron resonance gas model (HRG), since this model provided a good description for the thermodynamical evolution of the hadronic system below the critical temperature~\cite{Karsch:2003vd,Karsch:2003zq,Redlich:2004gp} and has been used to characterize the conditions deriving the chemical freeze-out~\cite{Taw3b,Taw3c}.

In many aspects, the fluctuations in particle production are essential, particularly with regard to examining the existing statistical models \cite{giorg,Karsch:2003vd},  characterize particle equilibration \cite{pequil} and search for unambiguous signals for the creation of {\it new state of matter} \cite{qgpA,qgpB}. 
Within the context of the statistical models of particle production \cite{statst}, it has been concluded that none of these models is perfectly able to describe the
experimental data. Using grand--canonical ensembles and quark phase space occupation factor $\gamma$ makes the comparison relatively better. In HRG with grand--canonical ensembles, the dynamical fluctuations of particle yield ratios, including kaon--to--pion, have been studied \cite{tawPS}. To bring theoretical and experimental data very close to each others, the factor $\gamma$ played an essential role. It has been noticed that the fluctuations over the whole range of the available center--of--mass energies $\sqrt{s}$ exhibit a non-monotonic behavior and $\gamma$ is varying with increasing $\sqrt{s}$, so that SPS-- are different from RHIC--data.     

In present work, we recall the concept of phase space dominance suggested by Fermi \cite{fermii}, six decades ago. The assumption of {\it equilibrium} single--particle distribution function $f$ and {extensive} thermodynamics that have been shown to perfectly reproduce all thermodynamic quantities, including multiplicities and  fluctuations, as long as the energy density is not high enough to derive the hadronic system into {\it new state of matter}, quark--gluon plasma (QGP) should be no longer valid, when the energy density exceeds a  critical value, as the case at RHIC and LHC energies. Across the phase transition, the symmetries and accordingly the effective degrees of freedom are likely subject of a prompt change. Therefore, the phase space volume, in which the microscopic states are distributed according to $f$ turns to a subject of a change as well. In present work, we implement the change of phase space volume and $f$ in the grand--canonical partition function, especially at large $\sqrt{s}$. It is not a theory describing $\sqrt{s}$--dependence of $f$ and/or the dynamics of changing phase space volume. This would be rather the role of QCD. For this purpose, Boltzmann equation integrated over momentum--space can alternatively be used. This has been applied on characterizing freeze-out of nucleosynthesis in the early universe \cite{nucleo}. 

Appart from these argumentations, this idea, from the phenomenological point--of--view, does not appear from nowhere. While studying particle number and ratio fluctuations, there have been speculations about the possibility of phase space change \cite{tawPS} with increasing $\sqrt{s}$. The entropy per particle, $S/N$, has been analyzed over the available range of $\sqrt{s}$. It was clear that $S/N$ is able to reflect a kind of rapid modification in phase space volume. The role of $\gamma$, the quark phase space occupation factor, can then be carried out by $S/N$. It has been concluded that the dependence of phase space volume on energy is likely essential to access the phase transition, at large $\sqrt{s}$. For kaon production in heavy--ion collisions, especially the horn at top SPS energy, an initial partonic phase has been assumed beyond this energy threshold \cite{kpi_horn}. Below it, a hadronic initial state has been utilized. The dependence of kaon--to--pion fluctuations on both energy-- and multiplicity--scaling has been analyzed, recently\cite{kochSchuster}, where the multiplicity of kaon-- and pion--yields has been scaled out depending on the dynamical fluctuations. These studies would explain why kaon--to--pion ratios are thought to be interesting. Furthermore, the measured minima at top SPS energy would be related to an increase of entropy production and a decrease of strangeness--to--entropy ratio. In heavy--ion collisions, the strangeness dynamical fluctuations are of great interest, since they are sensitive to the equation of state and microscopic structure of strongly interacting matter created at early stage of the collisions \cite{grons}. 

The paper is organized as follows. The model is given in section \ref{sec:model}, where single--particle equilibrium distributions of hadrons and partons are introduced. Section \ref{sec:dynmFlct} is devoted to the dynamical fluctuations of kaon--to--pion ratio. Discussion and final conclusions are elaborated in section \ref{sec:concls}.

\section{Model}
\label{sec:model}

\subsection{Single--Particle Equilibrium Distribution of Hadrons}

The grand-canonical partition function is given by Hamiltonian and baryon number operators, $\hat{H}$ and $\hat{b}$, respectively, and depends on temperature $T=1/\beta$ and chemical potential $\mu$, 
\bea \label{eq:zTr}
Z(\beta,V,\mu) &=& \Tr \, \left[\gamma\, \exp^{\beta (\mu \hat{b}-\hat{H})}\right].
\eea
It can be characterized by various but a complete set of microscopic
states and therefore the physical properties of the quantum systems turn to be obtainable in approximation of non-correlated {\it independent} hadron resonances. The resonances treated as a free gas~\cite{Karsch:2003vd,Karsch:2003zq,Redlich:2004gp,Tawfik:2004sw,Taw3} are conjectured to add to the thermodynamic pressure of hadronic matter. This statement is valid for free, as well as, strong interactions of the resonances. It has been shown that the thermodynamics of strongly interacting system can be approximated to an ideal gas composed of hadron resonances with masses $\le 2~$GeV~\cite{Tawfik:2004sw,Vunog}. The main motivation of using the Hamiltonian is that it contains all relevant degrees of freedom of confined and strongly interacting matter. It implicitly includes the interactions that result in formation of new resonances. In addition, this model has been shown to provide a quite satisfactory description of the particle production in heavy ion collisions. 

The distribution of resonances in the available states in micro-canonical, canonical (statistical) and grand-canonical ensemble is subject of statistical physics. Here, we consider the occupation number basis of single--particle \cite{raflsk} as the one which is suitable to evaluate Eq. (\ref{eq:zTr}). Each macro--state is to be characterized by $n\in\{n_i\}$ set of allowed {\it micro}states, their $b_i$ baryon charges and $\epsilon_i=(\vec{k}^2+ m_i^2)^{1/2}$ energies of i-th single state of mass $m_i$ and quantized longitudinal conjugate momenta $\vec{k}$. The effective mass $m_i=(m0_{i}^2+\vec{kt}_i^2)^{1/2}$, where $m0_{i}$ is the physical mass and $kt$ is the transverse momenta. In the relativistic limit $kt<<k$, the effective mass refers to the physical one. Obviously, the {\it macro}--state energy $E_i$ is given by the sum over all possible states, i.e, $\sum_in_i \epsilon_i$, where the number $n_i$ depends on whether the resonance is fermion or boson. For bosons, $n_i$ can be very large. For fermions, $n_i$ is subject of Pauli occupation principle. This summation counts all possible basis states. To make any physical treatment, one would need to make approximations  according to the required precision. In this regard, the summation can be replaced by integration over the controlling variable, for instance, over $\epsilon$, which in turn depends on $\vec{k}$, where $\vec{k}$ is subject of quantization. Therefore, the summation over $\epsilon$ is to be replaced by an integral over all $\vec{k}_i$ in the momentum space. In infinite volume\footnote{The volume of the space volume is given by $\int_0^{\infty}\Pi_i d^ix d^ik$, where $x$ is the coordinate. In many cases, the geometry is either simple or known and, therefore, symplectic manifolds can to be assumed, where a two-form $\omega=\omega_{ij} dx^i/2 \wedge dx^i$ is fulfilled. Symplectic forms are -- per definition -- closed and non-degenerate.} limit, $V\equiv dx^3$, the phase-space integral replaces the sum over all possible $2\times n$-dimensional phase-space volume element $d^3x\, d^3p$, which apparently gives the probability of finding the particle in this infinitesimal region of phase-space, i.e, $\sum_i=g/(2\pi)^3\int_0^{\infty} d^3x\,d^3p$. $n$ refers to the dimensions of the physical system. As given above, we used $n=3$. The phase space distribution extends mainly along the large longitudinal momenta. 

In equilibrium \cite{lagrangeBook}, the average energy $E=\sum_i n_i\epsilon_i$, and particle counts $N=\sum_i n_i$ of macroscopic states are conjectured to be know and remain constant. With respect to these two constrains, one can take the Hamiltonian on the system as an observable quantity. Here, we only consider the configuration which maximizes the number of {\it micro}states, instead of considering all configurations $\{n_i\}$. The maximum count, $\Omega=(N_i!/\Pi_i n_i!)\Pi_i g_i^{n_i}$, where $g_i$ is the degeneracy factor of the configuration $n_i$ of $i$--th microstate. It is related to the global entropy $S=\ln(\Omega)$ and information on the system compared to the measurements. This maximum is a formal subject to the external constraints on intensive variables $E$ and $N$. Since the states are conjectured not to be correlated with each other, the chemical potential $\mu$ is vanishing. Then, the maximum number of {\it micro}states is given by solving
\bea
\frac{\partial}{\partial n_j}\left(S-\alpha N -\beta E \right)=\frac{\partial}{\partial n_j}\left(\ln N! + \ln \Pi_i^n g_i^{n_i} - \sum_i^n \ln n_i! - \alpha N -\beta E \right) &=&0,
\eea
which means that only the terms having same subscript $j$ remain finite. The coefficients $\alpha$ and $\beta$ are Lagrange multipliers in entropy maximization. Each of these multipliers basically adds some unknown amount of each independent constraint to the function being optimized and ensures that the constraints are satisfied. 
\bea
\frac{\partial}{\partial n_j} \ln \Pi_j g_
j^{n_j}- \frac{\partial}{\partial n_j}\ln n_j! - \alpha -\beta \epsilon_j  &=&0.
\eea
Utilizing the Stirling approximation, then the occupation number,
\bea
n_j&=& g_j\,\exp\left(-\alpha-\beta \epsilon_j\right),
\eea
which apparently falls off exponentially with increasing $\epsilon$, since, as will be shown below, $\gamma=\exp(-\alpha)$ is constant.

The physical meaning of $\beta$ is, as given above, the inverse of temperature \cite{LM1}. Assuming two configurations $n_i$ and $n_j$, then $\beta=[\ln(n_i/n_j) + \ln(g_i/g_j)]/(\epsilon_i-\epsilon_j)$. Therefore, $\beta$ can alternatively be defined as the change in microscopic configurations per energy change. Apparently, this definition is consistent with the second law of thermodynamics. From thermodynamics, especially the identification of the Gibbs equation, $\alpha$ is another variable controlling the number of particles in the phase space.  
\bea
\alpha &=& \ln\left(\frac{\sum_i^n g_i\,\beta\epsilon_i}{N}\right)=\ln \sum_i^n g_i\, \beta\epsilon_i - \ln N.
\eea
Then, the occupation number reads
\bea
n_j &=& g_j\, \left(\frac{N}{\sum_i^n g_i\, \beta\epsilon_i}\right) \; \exp\left(-\beta\,\epsilon_j\right) = g_j\, \left(\frac{N\,T}{\cal E}\right) \; \exp\left(-\beta\,\epsilon_j\right),
\eea
where ${\cal E}=\sum_i^n g_i\epsilon_i$ is the summation of all possible single--particle state energies multiplied by the effective degrees of freedom. The multiplier $\alpha$ can be written as
\bea
\alpha &=& \ln \left(\frac{\cal E}{N\,T}\right) = \ln {\cal E} - \ln T - \ln N.
\eea
It combines intensive variables, $T$ and $N$ with an extensive one ${\cal E}$. 
The most probable state density is to be found by Lagrange multipliers, where one of them, $\alpha$, has been expressed in term of the second one, $\beta$, and the occupation numbers of the system. Apparently, $\alpha$ gives how the energy ${\cal E}$ is distributed in the microstates of the equilibrium system and therefore, can be understood as another factor controlling the number and thus is similar to chemical potential at the microcanonical level.

Besides the two constrains given previously, the conservation of baryon number $n_b$ represents an additional constrain on the grand--canonical partition, Eq. (\ref{eq:zTr}). Therefore, 
\bea \label{eq:zTr-Lagrange}
Z_{gc}(T,V,\mu) &=& \Tr \, \left[\exp^{\frac{\mu \hat{b}-\hat{H}}{T}-\alpha}\right],\\
f_{gc}(T,V,\mu) &=& \frac{\exp\left(\frac{-\hat{H}}{T}-\alpha\right)}{Z_{gc}(T,V,\mu)}.
\eea

With these assumptions, the dynamics of the partition function can be calculated as sum over {\it single--particle partition} functions $Z_{gc}^i$ of all hadrons and resonances.
\bea
\ln Z_{gc}(T,V,\mu)&=&\sum_i \ln Z_{gc}^i(T,V,\mu)=\sum_i\pm \frac{g_i}{2\pi^2}\,V\int_0^{\infty} k^2 dk \ln\left(1\pm \gamma\,\lambda_i\, e^{-\epsilon_i(k)/T}\right)
\eea
where $\pm$ stands for bosons and fermions, respectively.  $\lambda_i=\exp(\mu_i/T)$ is the $i$-th particle fugacity. $\gamma=\exp(-\alpha)$ is the quark phase space occupation factor.

\subsection{Single--Particle Equilibrium Distribution of Partons}
 
The Fokker-Planck equation is a well-known tool used to study the dynamics and velocity distribution of objects in thermal background, such as the transport properties in quark-gluon plasma \cite{2ppA,3ppBa,3ppBb}. Studying the stochastic behavior of a single object propagating with random noise, known as Langevin equation \cite{4pp1,7pp2}, represents one way to solve this problem. A master equation, such as the linearized Boltzmann-Vlasov equation \cite{6pp3}, with the Landau soft-scattering approximation would give another solving method.

As given previously, the statistical properties of an ensemble consisting of individual {\it parton} objects is given by a single--particle distribution function $f$. The probability of finding an object in infinitesimal region of the phase-space is directly proportional to the phase--space volume element and the distribution function describing it, where $f$ -- in this case -- is assumed to fulfill the Boltzmann-Vlasov (BV) master equation, which is the semi--classical limit of a time--dependent Hartee--Fock theory through the Wigner transform of the one--body density matrix and can be used to study the dynamics of the constituents quarks in hadrons,
\bea \label{eq:te1}
\dot f + \dot{\vec{x}} \cdot \nabla_x f + \dot{\vec{k}} \cdot \nabla_k f + \dot{\vec{q_c}} \cdot \nabla_{q_c} f &=& {\cal G}  + {\cal L}.
\eea
In rhs, the first term ${\cal G}$ represents gain, i.e, the rate that a particle with momentum $k + kt$ loses momentum $kt$ due to reactions with the background and the second term ${\cal L}$ represents loss due to the scattering rate. Therefore, these two terms furnish us with details about the interaction. Details about participating partons and their reactions have been discussed in literature, for example in Ref. \cite{muellr1}. Seeking for simplicity, a two--body system has been assumed. Then the effective potential $U$ has to combine the well--known Coulomb  $U(x\rightarrow 0)\propto 1/x$ and confined potentials $U(x\rightarrow \infty)\propto 0$. As assumed, the particles of interest are partons currying electric and color charges. In lhs, standard position $\vec{x}$ and momentum $\vec{k}$ variables are given in the first two terms. The third term  represents the dynamics of the charge, where $\dot{\vec{k}}$ can be given by field tensor (Lorentz relativistic force). The fourth term reflects an extension of phase space to include color charge, $\dot{\vec{q_c}}$. 
  
One of the largest advantages of the transport equation, Eq. (\ref{eq:te1}), is that the phase space distribution functions of partons and/or hadron resonances by test--particle distributions for these different species. In test--particle approach \cite{wanng}, the continuous distribution is discritized to a finite number of test--particles representing the individual phase space cells. The test--particle itself obeys Newtonian equation--of--motion \cite{BALi91}. Therefore,  
\bea
\frac{d \vec{x}}{d t} &=&  \frac{\vec{p}}{E}, \\
\frac{d \vec{k}}{d t} &=& \vec{F}(\vec{x})+\cdots,\\
\frac{d \vec{q_c}}{d t} &=& f^{abc} u_{\mu} q_b A^{c \mu},
\eea
where $q_c$ represents the chromofield exchange of quarks and gluons in the presence of the background field. $\vec{F}$ can be expressed, as mentioned above, in Lorentz force, $q_c(\vec{E}+\vec{v}\times\vec{B})$ or any effective potential and higher terms can be replaced by the rest of the Langevin system of equations.
As given in Ref. \cite{muellr1,PonaserA}, BV master equation is solvable for  
\bea\label{fHadron1}
f(\vec{x}, \vec{k}, t) &=& \frac{1}{n_{tp}} \sum_i^N g_x(\vec{x}-\vec{x}_i(t)) \; g_k(\vec{k}-\vec{k}_i(t)),
\eea
where $n_{tp}=N/q_c$ is the total number of test--particles per quark charge. In this model, the test--particle can be the hadron yields, in which we are interested. The functions $g_x$ and $g_k$ can be Heaviside or any sharply peaked distribution function \cite{PonaserA}. In order to make the physical measurements possible, they have to be modeled \cite{muellr1}, for example as Gaussian-type, 
\bea
g_x(\vec{x}-\vec{x}_i(t))&=&\sqrt{\frac{\omega}{\pi}} \exp\left\{-\omega\left[\vec{x}-\vec{x}_i(t)\right]\right\},
\eea
where $\omega=1/2\sigma^2$ and the invariance $\sigma>0$. Positions $\vec{x}_i$ and momenta $\vec{k}_i$ vary around $\vec{x}$ and $\vec{k}$, respectively. Eq. (\ref{fHadron1}) can be interpreted as a measure for the deviation from equilibrium or the relaxation toward equilibrium.   

The Hamiltonian equation--of--motion of the test--particle is obtainable by inserting Eq. (\ref{fHadron1}) in Eq. (\ref{eq:te1}). The function $f$ reaches equilibrium, when the resulting probability current vanishes \cite{3ppBb,crrent}. To illustrate this, let us recall the Tsallis distribution \cite{isaliis} depending on the parameter $q$, which measures the degree of extensivity of the entropy $S$ in the system. Then the single--particle distribution function reads
\bea \label{eq_sTsalis1}
f(\vec{x}, \vec{k}) &=& N \left[1-\frac{\epsilon(\vec{k})}{T}(1-q)\right]^{1/(1-q)},
\eea
which obviously gives the {\it extensive} Boltzmann limit, when $q\rightarrow 1$. In Eq. (\ref{fHadron1}), the Boltzmann limit is reachable at very large time $t$, i.e, when the number density turns to be conserved or the change in entropy $S$  vanishes. Thus, the number density and distribution function would be written as
\bea
n(\vec{x},\vec{k},t) &=& n_{eq}(\vec{x},\vec{k}) + f(\vec{x},\vec{k},t), \label{eq:nn}\\
f(\vec{x},\vec{k},t) &\simeq& f_{eq}(\vec{x},\vec{k})\; {\cal Q}(\vec{x},\vec{k}), \label{eq:nf}
\eea
where ${\cal Q}(\vec{x},\vec{k})\simeq\exp(2\omega\,\vec{x}\,\vec{k})$ exponentially raises with increasing position $\vec{x}$ and momentum $\vec{k}$ of the test--particle. It reflects the change in the phase--space when the hadronic degrees of freedom are replaced by the partonic ones. The factor ${\cal Q}$ would not be too much different than $q$. Therefore, it can be interpreted as a measure for the non--extensivity. The exact relation between $q$ and ${\cal Q}$, especially in the context of heavy--ion collisions \cite{TsallisHIC}, will be the subject of an upcoming work. $f_{eq}(\vec{x},\vec{k})$ is an equilibrium normalized distribution function similar as the one assumed and utilized in previous subsection. In deriving Eq. (\ref{eq:nf}), we assumed that the products of $\vec{x}\,\vec{x}_i$ and $\vec{k}\,\vec{k}_i$ vanish, where $\vec{x}_i$ ($\vec{k}_i$) is position (momentum) of the i-th cell in the phase space volume.

Non-equilibrium processes can be originated from different sources. Previously, we assumed that the phase transition is a dominant source. Because of presence of non–-Markovian processes in the kinetic equation, the extreme conditions in ultra--relativistic heavy-ion collisions are conjectured to produce long–-range color interactions and so--called memory effects. They likely affect the standard equilibrium distributions and even the thermalization processes toward the  equilibrium state \cite{colr1}. We therefore suggested in Eq. (\ref{eq_sTsalis1}) that the distribution function is to expressed out in Tsallis statistics. Based on distributions of partons in the microstates of the phase--space volume, a rigorous determination of non--extensivity is given in Eq. (\ref{eq:nf}). Therefore, the single--particle distribution function reads in the grand--canonical ensemble,
\bea
f_{gc}(\vec{k}, \mu) &=& \left\{\left[1+\left(\frac{\epsilon(\vec{k})-\mu}{T}-\alpha\right)(q-1)\right]^{1/(q-1)}\pm 1\right\}^{-1}, \label{eq:nfinal1}\\
   &=& \left\{ \gamma^{-1}{\cal Q}^{-1}(\vec{k})\exp\left[\frac{\epsilon(\vec{k})-\mu}{T}\right] \pm 1\right\}^{-1},\label{eq:nfinal2}
\eea
where in Eq. (\ref{eq:nfinal2}), the volume compensates the corresponding part of ${\cal Q}(\vec{x},\vec{k})$. This procedure gives solid reasons not only for the non--extensivity, but also for the assumed modification in size of system and phase space volume. Then, the grand-canonical partition function can be written as  
\bea 
\label{eq:zTr2Tsl}
\ln Z_{gc}(T,V,\mu)&=& \sum_i\pm \frac{g_i}{2\pi^2}\,V\int_0^{\infty} k^2 dk\,\ln\,\left(1\pm \left[1+\left(\frac{\epsilon_i(\vec{k})-\mu_i}{T}-\alpha\right)(q-1)\right]^{1/(q-1)}\right), \\
\label{eq:zTr2}
\ln Z_{gc}(T,{\cal V},\mu)&=& \sum_i\pm \frac{g_i}{2\pi^2}\,{\cal V}\int_0^{\infty} k^2 dk\; \ln\,\left(1\pm \gamma\,{\cal Q}\, \exp\left[\frac{\mu_i-\epsilon_i(k)}{T}\right]\right),
\eea
where the pressure in this grand-canonical ensemble is given as 
\bea
p_{gc}(T,{\cal V},\mu)&=& \lim_{{\cal V}\rightarrow \infty} \frac{T}{{\cal V}} \ln Z_{gc}(T,{\cal V},\mu). 
\eea

\section{Dynamical Fluctuations in Kaon--to--Pion Ratios}
\label{sec:dynmFlct}

The fluctuations in particle number are given by the susceptibility, which is the derivative of particle number $\langle n\rangle$ wrt chemical potential $\mu$.  
\bea 
\label{eq:n1} 
\langle n\rangle &=& \sum_i \frac{g_i}{2\pi^2} \int_0^{\infty} k^2 dk 
\frac{\gamma\,{\cal Q}}{\exp\left[\frac{\epsilon_i(k)-\mu_i}{T}\right] \pm \gamma\,{\cal Q}}, \\
\label{eq:dn1} 
\langle (\Delta n)^2\rangle &=& \sum_i \frac{g_i}{2\pi^2} \frac{1}{T} \int_0^{\infty} k^2 dk 
           \frac{\gamma\,{\cal Q}\,\exp\left[\frac{\epsilon_i(k)-\mu_i}{T}\right]}
	   {\left(\exp\left[\frac{\epsilon_i(k)-\mu_i}{T}\right] \pm \gamma\,{\cal Q}\right)^2}.
\eea

When chemically relaxing system absolves the chemical freeze-out process, the hadron resonances are conjectured to decay either to stable particles or
to other resonances. This chemical process has to be take into account in the particle numbers and fluctuations as follows.
\bea \label{eq:n2}
\langle n_i^{final}\rangle &=& \langle n_i^{direct}\rangle + \sum_{j\neq
i} b_{j\rightarrow i} \langle n_j\rangle,\\ \label{eq:dn2} 
\langle (\Delta n_{j\rightarrow i})^2\rangle &=& b_{j\rightarrow i} (1-b_{j\rightarrow
i}) \langle n_j\rangle + b_{j\rightarrow i}^2 \langle (\Delta
n_{j})^2\rangle 
\eea
where $b_{j\rightarrow i}$ being branching ratio for the decay of $j$-th resonance
to $i$-th particle. In order to characterize the stage at which the chemical freeze--out takes place, we assume that the ratio $s/T^3$, where $s$ is the entropy density, gets a constant value \cite{Taw3b,Taw3c,Taw3}.

For the event--by--event fluctuations of the ratio of two particles $K/\pi$ (kaon--to--pion) are~\cite{koch}
\bea  \label{eq:sigma}
\sigma^2_{n_K/n_{\pi}} &=& \frac{\langle (\Delta n_K)^2\rangle}{\langle n_K\rangle^2} + 
                       \frac{\langle (\Delta n_{\pi})^2\rangle}{\langle n_{\pi}\rangle^2} - 
                     2 \frac{\langle \Delta n_K \; \Delta
		     n_{\pi}\rangle}{\langle n_K\rangle \; \langle
		     n_{\pi}\rangle},
\eea
which combine dynamical and statistical fluctuations. Third term of
Eq.~\ref{eq:sigma} counts for fluctuations from the hadron resonances which 
decay into $K$ and $\pi$, simultaneously. In such a
mixing channel, all correlations including quantum statistics ones are
taken into account.  Obviously, 
this decay channel results in strong correlated particles. To extract
statistical fluctuation, we apply Poisson scaling in mixed decay
channels. Experimentally, there are various methods to
construct statistical fluctuations~\cite{STAR}. Frequently used
method is the one that measures particle ratios from mixing events.
\bea \label{eq:sigmaStat}
(\sigma^2_{n_K/n_{\pi}})_{stat} &=& \frac{1}{\langle n_K\rangle} +
\frac{1}{\langle n_{\pi}\rangle} 
\eea
Subtracting Eq.~\ref{eq:sigmaStat} from Eq.~\ref{eq:sigma}, we get
dynamical fluctuations of $n_K/n_{\pi}$ ratio. 
\bea
 \label{eq:sigma2}
(\sigma^2_{n_K/n_{\pi}})_{dyn} &=& 
          \frac{\langle n_K^2\rangle}{\langle n_K\rangle^2} +
          \frac{\langle n_{\pi}^2\rangle}{\langle n_{\pi}\rangle^2} -
         \frac{\langle n_K\rangle+\langle n_{\pi}\rangle +
	 2\langle n_K n_{\pi}\rangle}{\langle n_K\rangle\langle n_{\pi}\rangle}
\eea

\begin{figure}[thb]
\includegraphics[height=8.cm]{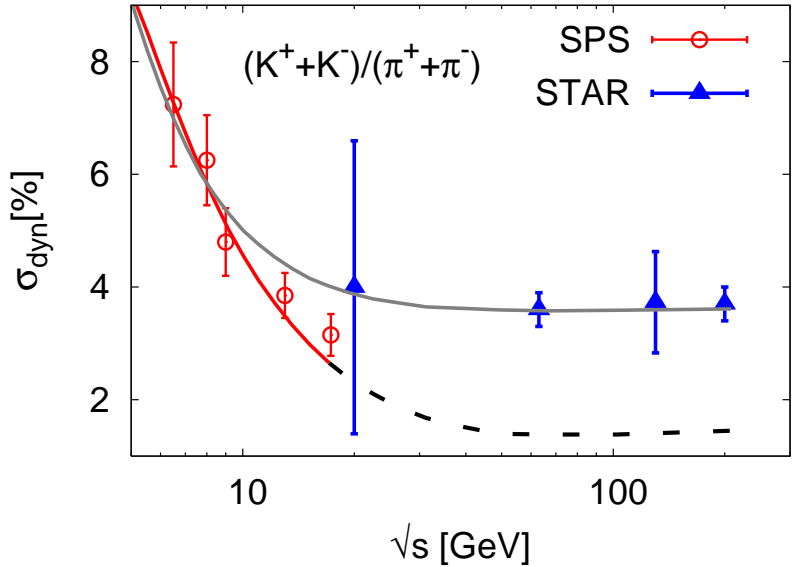}
\caption{\normalsize Dynamical fluctuations of $(K^++K^-)/(\pi^++\pi^-)$ ratio as function of center--of--mass energy $\sqrt{s}$. Taking into account the experimental acceptance, the lower curve represents HRG results at finite value for $\gamma$. It reproduces perfectly the SPS data. The dashed region shows that RHIC data are largely underestimated. The upper curve represents HRG results, when phase space factor ${\cal Q}$ is switched on. RHIC data perfectly matches with this curve.} 
\label{fig:1a}
\end{figure}

Present model on the dynamical fluctuations and their modifications with varying phase space volume, especially when the center--of--mass energy is high enough to secure energy density able to derive the hadronic mater into QGP, this model is valid for any particle yield rations. The dynamical fluctuations in kaon--to--pion ratio, $(K^++K^-)/(\pi^++\pi^-)$, are given in Fig. \ref{fig:1a}, since they combine strangeness and light--boson fluctuations. They are very sensitive to deconfinement and chiral phase transitions, respectively.  

As introduced in previous subsections, the HRG model assumes that the hadron resonances are point-like and non--correlated. The dynamical fluctuations of particle yield ratios with and without ${\cal Q}$ are produced and plotted in Fig. \ref{fig:1a}. It is obvious that the average multiplicity $\langle n\rangle$, Eq.~(\ref{eq:n2}), and the dynamical fluctuations $\sigma^2_{N_1/N_2}$, Eq.~(\ref{eq:sigma}), are not strongly dependent on the volume fluctuations. Therefore, we assume -- in this work -- that volume fluctuations over the entire range of $\sqrt{s}$ are minimum and thus neglected. The experimental acceptances of the different detectors have been taken into account. Also, the quark phase space occupation factor, $\gamma$, has been estimated. So far, we conclude that HRG reproduces the experimentally measure fluctuations of $(K^++K^-)/(\pi^++\pi^-)$ ratio over the entire range of $\sqrt{s}$. In generating this excellent agreement no fitting has been performed. The comparison is given in Fig. \ref{fig:1a}.

\begin{figure}[thb]
\includegraphics[height=8.cm]{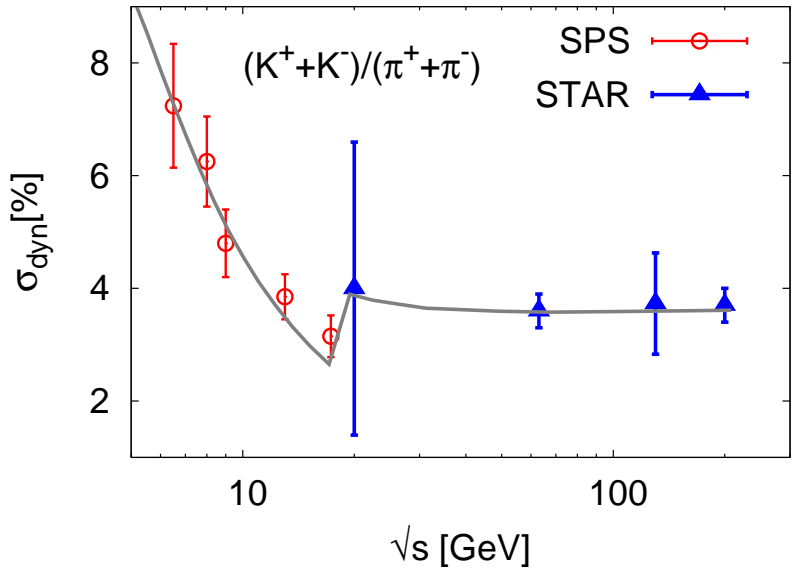}
\caption{\normalsize Same as in Fig. \ref{fig:1a}. Implementing the experimental acceptances of the different detectors, keeping $\gamma$, that exclusively has been used to produce the curve below $\sqrt{s}\sim 17$ GeV, and utilizing phase space factor ${\cal Q}$ at higher $\sqrt{s}$, result in the curve that obviously very well describes the entire experimental data set. } 
\label{fig:2a}
\end{figure}

\section{Discussion and Conclusions}
\label{sec:concls}

The scale invariant processes are conjectured to lead to intermittency. i.e, a power--law dependence of factorial moments $F$ on the phase space volume, $F^{(k)}(\delta {\cal V})\propto \delta {\cal V}^{-\delta_k}$, where $\delta {\cal V}\simeq\delta \vec{x}\delta \vec{k}$ is the phase space volume element and $\delta_k$ is known as the intermittency exponent \cite{bialas1}. The importance of phase space dimension in the intermittency approach has been discussed by Ochs \cite{ochs}. Fluctuations and correlations in multi--particle distribution are accessible through intermittency. Also, information about dynamics of the decaying system would be available by this statistical concept that includes self-similarity or fractality of hadron production \cite{intrmtAgAg}. Also the parton shower would exhibit intermittency. Nevertheless, there is no guarantee that it survives until the hadronic final state \cite{fractall}. In framework of Ginzburg-Landau theory, the intermittency in phase transition, hadrons--QGP, has been studied \cite{bab1}. The intermittency evolution has been bound with the dynamics of QCD critical point, where it was concluded that the freeze-out profile would present a structure revealing traces of critical fluctuations \cite{antoniou}. 

Using generalization of central limit theorem applied to random cascading models, 
Levy stable laws have been introduced to describe and classify the intermittency patterns due to different kinds of phase transitions, for instance hadrons--QGP \cite{levyBrax,levyTawfik,gang}. In previous work, we studied the one-- and two--dimensional intermittency in Pb--Pb collisions at $158$ AGeV/c, i.e, top SPS energy \cite{levyTawfik}. The exponents, $\alpha_k$, are compared with the anomalous dimensions, $q$, which has been expressed in terms of Levy stable index. It has been found that the fluctuations within a narrow rapidity internal fulfill the requirements of the Levy stability, referring to phase transition, i.e, positive Levy stable index. In the same year, an interpretation of the non--extensitivity parameter, $q$, in Levy and Tsallis distributions has been suggested \cite{levy-dq}.

This discussion aims to connect the event--by--event dynamical fluctuations with the intermittency and refer to earlier phenomenological results that top SPS energy has been enough to secure energy density able to dissolve the hadronic matter into QGP. The non--extensitivity given above in the anomalous parameter $q$ and previously in Eq. (\ref{eq:nfinal1}) is connected with departure from equilibrium and therefore has to be taken into account, when the system is assume to go through phase transition, as the case at top SPS energy and beyond.   

As shown in Fig. \ref{fig:1a}, the dynamical fluctuations are compared with the HRG results. The quark phase space occupation factor $\gamma$ is apparently not able to reproduce the whole data sent, although the experimental acceptances of the different detectors have been taken into considerations. In addition to the assumptions introduced in \cite{kpi_horn,kochSchuster}, we given a novel one. We assume that the prompt raise at $\sqrt{s}\sim17$ GeV is to be understood according to modification in the phase space volume. To this end, we studied the single--particle distribution functions $f$ of hadrons and partons. It has been concluded that $f$ is a subject of modification, especially at large $\sqrt{s}$. In this limit, the energy density available to the system turns to be high enough to cause the hadronic matter, where equilibrium $f$ is perfectly able to reproduce almost all essential transport properties and thermodynamic quantities, including the dynamical fluctuations, to go through a phase transition to QGP. Such a phase transition apparently results in various types of modifications, such as symmetries and degrees of freedom. Also the configurations of microstates in the phase space volume $d^3\vec{x}d^3\vec{k}$ and the single--particle distribution function $f$ are not exceptions.   
 
We apply this model, exclusively, to the dynamical fluctuations of $(K^++K^-)/(\pi^++\pi^-)$ ratios, since they combine both strangeness and light boson fluctuations and have one of the largest experimental acceptances in the different detectors. Additionally, they are very sensitive to the deconfinement and chiral phase transitions, respectively. Furthermore, the strangeness dynamical fluctuations are expected to survive through the mixed phase. Implementing $\gamma$ and ${\cal Q}$ in the grand canonical partition function of HRG results in a very well description of the experimentally measured fluctuations of these particle yields over the entire range of $\sqrt{s}$. The comparison is drawn in Fig. \ref{fig:2a}. The non-monotonic behavior is very well reproduced over a wide range of $\sqrt{s}$. It is obvious, that these dynamical fluctuations non--avoidablely refer to non-extensive and non--equilibrium state of matter, that basically differs from the one at SPS energies. The modification of the state of matter has been combined in the factor ${\cal Q}(\vec{x},\vec{k})\simeq\exp(2\omega\,\vec{x}\,\vec{k})$ whose numerical value is roughly estimated and kept unchanged with changing $\sqrt{s}$.    

%-----------------------------------------------

%-----------------------------------------------

\end{document}